\documentclass[conference]{IEEEtran}
\IEEEoverridecommandlockouts
\usepackage{cite}
\usepackage{amsmath,amssymb,amsfonts}
\usepackage{algorithmic}
\usepackage{graphicx}
\usepackage{textcomp}
\usepackage{xcolor}
\usepackage{threeparttable}
\usepackage{array}
\usepackage{dblfloatfix}
\usepackage{graphicx}

\def\BibTeX{{\rm B\kern-.05em{\sc i\kern-.025em b}\kern-.08em
    T\kern-.1667em\lower.7ex\hbox{E}\kern-.125emX}}
\begin{document}

\title{AmpLyze: A Deep Learning Model for Predicting the Hemolytic Concentration\\
    \thanks{*Equal contribution. †Corresponding author.}
}

\author{\IEEEauthorblockN{Peng Qiu$^{*}$}
\IEEEauthorblockA{\textit{School of Computer Science} \\
\textit{Carnegie Mellon University}\\
    Pittsburgh, PA 15213, USA \\
 pengq@andrew.cmu.edu}
\and
\IEEEauthorblockN{Hanqi Feng$^{*}$}
\IEEEauthorblockA{\textit{School of Computer Science} \\
\textit{Carnegie Mellon University}\\
Pittsburgh, PA 15213, USA \\
hanqif@andrew.cmu.edu}
\and
\IEEEauthorblockN{Meng-Chun Zhang$^{*}$}
\IEEEauthorblockA{\textit{Department of Biostatistics} \\
\textit{University of Pittsburgh}\\
Pittsburgh, PA 15261, USA \\
MEZ141@pitt.edu}
\and
\IEEEauthorblockN{Barnabas Poczos$^{\dagger}$}
\IEEEauthorblockA{\textit{School of Computer Science} \\
\textit{Carnegie Mellon University}\\
Pittsburgh, PA 15213, USA \\
bapoczos@cs.cmu.edu}
}

\maketitle

\begin{abstract}
In antimicrobial peptide development, red‑blood‑cell lysis (HC\textsubscript{50}) is the principal safety barrier, but existing in silico tools stop at a binary toxicity classification. Here we propose a new method, AmpLyze, that closes this gap by predicting the actual HC\textsubscript{50} value from protein sequence alone and explaining the residues that drive toxicity. The model couples residue‑level ProtT5/ESM2 embeddings with sequence‑level descriptors in dual local and global branches, aligned by a cross‑attention module and trained with log‑cosh loss for robustness to assay noise. The optimal AmpLyze model reaches a PCC of 0.756 and an MSE of 0.987, outperforming classical regressors and the state‑of‑the‑art. Ablations confirm that both branches are essential, and cross‑attention adds a further 1\% PCC and 3\% MSE improvement. Expected‑Gradients attributions reveal known toxicity hotspots and suggest safer substitutions. By turning hemolysis assessment into a quantitative, sequence‑based, and interpretable prediction, AmpLyze facilitates AMP design and offers a practical tool for early‑stage toxicity screening.
\end{abstract}

\begin{IEEEkeywords}
Hemolysis Prediction, Model Interpretability, Peptide Engineering
\end{IEEEkeywords}

\section{Introduction}
Escalating antimicrobial resistance since the widespread use of antibiotics has undermined many antibiotics’ efficacy \cite{cdc2019antibiotic}, driving the search for novel anti-infective approaches. Antimicrobial peptides (AMPs) are short and typically cationic molecules that form a first-line defense in the innate immune system. They have recently emerged as compelling candidates \cite{mba2022antimicrobial} for anti-infective approaches. They kill a broad range of microorganisms by preferentially disrupting microbial membranes or interfering with essential intracellular targets, and this biophysical mode of action places a comparatively low evolutionary pressure on pathogens to develop resistance \cite{bucataru2024antimicrobial}. Several candidates are already in clinical trials; notably, the magainin analogue pexiganan advanced to Phase III trials as a topical treatment for mildly infected diabetic foot ulcers \cite{lamb1998pexiganan}.   

However, conventional screening of AMPs is resource-consuming and labor-intensive, requiring extensive library design and numerous assays. Over recent decades, considerable research effort has focused on developing in silico methods for AMP discovery and optimization. Early computational approaches aimed at classifying peptides into non-AMPs and AMPs. Initial machine learning (ML) predictors such as AntiBP \cite{AntiBP}, AMPScanner vr.1 \cite{AMPScannerVr1}, and AMPfun \cite{AMPfun} primarily employed algorithms including Support Vector Machines and Random Forests on sequence-derived features like one-hot encodings of amino acids and hand-crafted protein descriptors such as hydrophobicity and net charge. The advent of deep learning models eliminated the need for tedious feature handcrafting and selection. In 2018, newer version AMPScanner vr.2 \cite{AMPScannerVr2} was the first attempt to apply a deep neural network to AMP classification. Subsequent deep-learning models such as AI4AMP \cite{AI4AMP}  and iAMPCN \cite{iAMPCN} built on this initial work. More recently, instead of relying solely on hand-crafted protein descriptors, advanced deep learning models like PepNet \cite{pepnet} have incorporated embeddings from large pre-trained protein language models to capture rich representations of long-range dependencies, structural motifs, and evolutionary signals. Notably, AMP-Identifier \cite{AMP-Identifier} adapted the pre-trained ProtBERT architecture, achieving an impressive 99.18\% accuracy in AMP classification.

Beyond basic classification, the field has progressed toward predicting quantitative potency metrics such as the minimum inhibitory concentration (MIC), which is the lowest concentration that stops microbial growth. Accurate MIC prediction can further narrow down promising candidates before costly assays. Yan et al. (2023) developed a multi‑branch CNN‑attention model for \emph{E.\ coli} \cite{yan2023deep}. Chung et al. then combined sequence features with protein‑language‑model embeddings using an ensemble method to improve MIC prediction for \emph{E. coli}, \emph{S. aureus} and \emph{P. aeruginosa} \cite{chung2024ensemble}. Most recently, Cai et al. applied BERT‑based transfer learning that further improved the Pearson correlation coefficient for \emph{E.\ coli} to around 0.8 \cite{cai2025bert}. These advances in potency prediction move beyond binary activity classification toward models that can guide dosage and efficacy considerations.

While the modeling of antimicrobial efficacy is advancing, an equally critical aspect of AMP development is toxicity, particularly hemolytic toxicity. Hemolysis is a primary indicator of an AMP's safety, which is usually measured as hemolytic concentration (HC\textsubscript{50}) representing the concentration of peptide required to lyse 50\% of red blood cells under defined assay conditions \cite{ruiz2014analysis}. A peptide, even with great potential in antimicrobial activity, will have a narrow therapeutic application if it lyses red blood cells at low concentration. Unfortunately, many potent AMPs, estimated at 70\% of known AMPs, also exhibit toxicity toward mammalian cells, particularly red blood cells, leading to hemolysis due to their amphipathic nature and high hydrophobicity \cite{plisson2020machine}. 

Only limited studies have been conducted from this toxicity perspective, and almost all of them focus primarily on binary classification of hemolytic and non-hemolytic AMPs. An early example is HemoPI \cite{HemoPI} that used multiple machine learning methods using amino acid composition features. Using the same dataset, HemoPred \cite{HemoPred} employed Random Forest to achieve better accuracy. More advanced techniques have since been explored. AMPDeep (2022) \cite{AMPDeep} used transfer learning by fine-tuning large protein language models to capture the complex sequence patterns associated with hemolytic activity. Similarly, ToxIBTL \cite{ToxIBTL} fused transfer learning with an information bottleneck to learn compact, toxicity-specific embeddings for enhanced peptide toxicity prediction. Recently, tAMPer (2024) \cite{tAMPer} introduced a multimodal architecture combining ESM2-derived embeddings and graph neural networks on predicted peptide structures, taking advantage of both sequence and structural information to make more accurate toxicity predictions.

Until very recently, only very few methods existed to predict the hemolytic concentration of a peptide. From a drug optimization and development perspective, a binary decision whether a candidate AMP is hemolytic or not is usually insufficient as there are substantial differences between a highly toxic peptide and one that's just above the hemolytic threshold. A predictor of hemolytic concentration can effectively guide the drug development process by allowing scientists to precisely tune peptide sequences. The most significant advance to date in this area came in 2025, where Rathore et al. trained a random forest regressor on Pfeature-derived physicochemical features achieving a Pearson correlation of 0.739 ($R^2$ = 0.543) on an independent test set \cite{rathore2025prediction}. 

Despite this progress, significant challenges remain in the accurate prediction of HC\textsubscript{50} due to limited size of available data and its inherently noisy nature. Experimental HC\textsubscript{50} determinations are sensitive to multiple assay conditions that complicate model training \cite{diff, greco2020correlation}. Furthermore, the relationship between sequence and hemolytic activity is highly non-linear, where even single amino acid substitutions can dramatically alter the toxicity through complex effects on peptide folding, aggregation, and membrane interactions \cite{kumar2016single, pirtskhalava2021physicochemical}. There is also an urgent need for model interpretability such that the importance of each amino acid position can guide the AMP optimization process. 

To address these challenges, we present AmpLyze, the first end-to-end deep learning model for HC\textsubscript{50} prediction. Our approach achieves state-of-the-art prediction performance while providing residue-level insights into the determinants of hemolytic toxicity. We evaluated AmpLyze using stratified 5‑fold cross‑validation. In each fold, one subset was held out as the test set, and the remaining data were split into training and validation sets. The model was trained on the training set and its hyperparameters tuned on the validation set before final evaluation on the test fold. Average performance across all five folds is reported.

\begin{figure*}[h]
    \centering
    \includegraphics[width=1\textwidth,page=1]{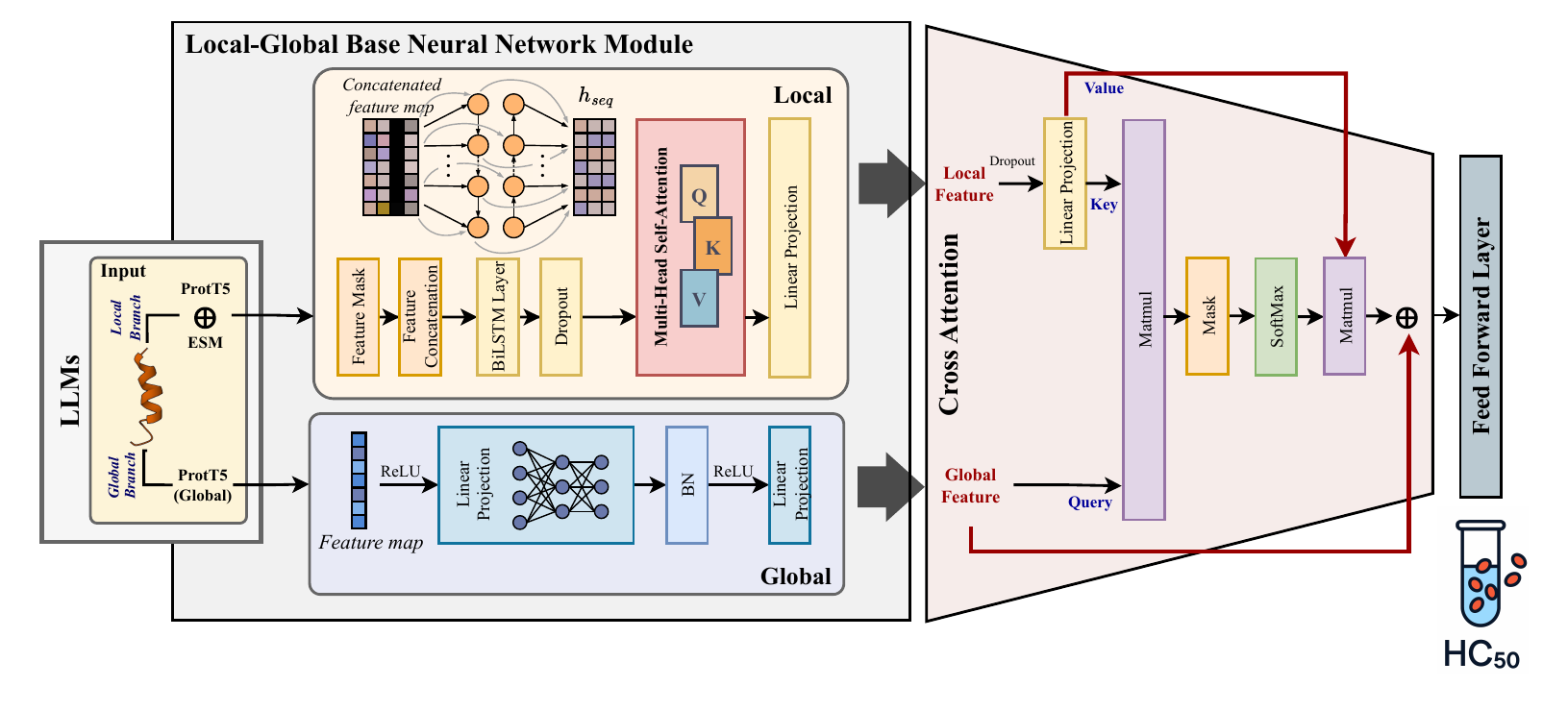}
    \caption{The proposed AmpLyze architecture based on local and global encoding branches and enhanced with a cross-attention fusion module.}
    \label{fig:model}
\end{figure*}

\section{PROPOSED METHODOLOGY}
Our architecture builds on the insight that large, self-supervised protein language models (pLMs) can learn rich biophysical and evolutionary context directly from sequence corpora, greatly reducing the need for handcrafted features. In this work, we employed two state-of-the-art pre-trained pLMs, ESM2-3B \cite{ESM2} and ProtT5-XL-UniRef50 \cite{ProtT5}, to encode the raw amino acid sequences. AmpLyze's dual-branch architecture extracts fine-grained residue embeddings from ESM2-3B alongside coarse-grained sequence embeddings from ProtT5-XL-UniRef50, then fuses these complementary representations dynamically through a learnable cross-attention module. By jointly modeling detailed local residue context and global sequence semantics, AmpLyze captures a richer biochemical embedding than either model alone.

\subsection{Local Encoding}
For each amino acid position $i$, we first extract two per-residue embeddings from pre-trained protein language models: one from ProtT5-XL-UniRef50 and the other from ESM2-3B, and then concatenate them into 
\begin{equation}
    r_i \;=\; \bigl[e_i^{\mathrm{T5}};\,e_i^{\mathrm{ESM}}\bigr]
\;\in\;\mathbb{R}^d.
\end{equation}
During training random feature masking is applied, randomly zeroing out channels of $r_i$ via an element-wise mask $m$. The masked vectors are projected into a lower-dimensional space and fed into a bidirectional LSTM, which integrates information from both N-terminus→C-terminus and C-terminus→N-terminus directions. Finally, a multi-head self-attention layer refines the sequence of LSTM outputs, yielding a contextualized representation matrix

\begin{equation}
\begin{split}
H_{\mathrm{seq}}
&= \mathrm{MHSA}\Bigl(
      \mathrm{BiLSTM}\bigl(
        W_{p}\,(m \odot [\,e^{\mathrm{T5}};\,e^{\mathrm{ESM}}\,])
        + b_{p}
      \bigr)
    \Bigr) \\
&\quad\in \mathbb{R}^{L \times d_{\mathrm{local}}}\, 
\end{split}
\end{equation}
where $L$ is the peptide length. 

\subsection{Global Encoding}
In parallel, we take the per-sequence ProtT5-XL-UniRef50 embedding, which is a single 1,024-dim vector summarizing the entire peptide, and pass it through a small two-layer MLP, giving the processed feature map $ g \in \mathbb{R}^{d_{\mathrm{model}}}$.

\subsection{Cross-attention Fusion}
To let the model decide which residues matter most given the overall peptide context, we use a cross-attention block where the global vector $g$ serves as the Query, and the per-residue features $H_{\mathrm{seq}}$ serve as both Keys and Values. Concretely, 
Key $\mathbf{K} = H_{\mathrm{seq}}\,W^{K}, $ and value $\mathbf{V} = H_{\mathrm{seq}}\,W^{V}$ are constructed with learnable weights $\mathbf{K}, \mathbf{V} \in \mathbb{R}^{d_{\mathrm{local}} \times d_{\mathrm{model}}}$. 
Attention scores are then computed as 
\begin{equation}
\boldsymbol\alpha 
\;=\;
\mathrm{softmax}\!\Bigl(\frac{\mathbf{q}\,\mathbf{K}^\top}{\sqrt{d_{\mathrm{model}}}} + M\Bigr)
\quad(\boldsymbol\alpha \in \mathbb{R}^{1\times L})\,
\end{equation} 
where $M$ is mask to prevent the model attending to padded positions. The attended residue summary $Y$ is computed as
\begin{equation}
    X = \boldsymbol{\alpha}\,V \;\in\;\mathbb{R}^{1\times d_{\mathrm{model}}}, 
\quad
Y = g + X \;\in\;\mathbb{R}^{1\times d_{\mathrm{model}}}.
\end{equation}

A scalar length feature is appended to $Y$. This vector is then passed through a fully connected layer to predict the hemolytic concentration. 

\section{EXPERIMENTS}

\subsection{Setup}
\paragraph{Dataset} We employed the same dataset as HemoPI2 to ensure a fair comparison of model performance. All peptide sequences were obtained from two publicly available sources: Hemolytik \cite{hemolytik} and DBAASP v3 \cite{dbaasp}. Hemolytik is a comprehensive resource that aggregates experimentally validated hemolytic and non-hemolytic peptides from several well-established antimicrobial peptide (AMP) databases, including APD2 and DAMPD, while DBAASP also contains detailed activity profiles that includes hemolytic and cytotoxic activity. Only AMP entries with available and numeric HC\textsubscript{50} values were retained; for peptides with multiple HC\textsubscript{50} measurements or reported ranges, the mean value was used to represent overall hemolytic activity. Sequences containing non‑standard residues or shorter than six amino acids were removed. After filtering, the final dataset comprised 1926 peptides. 

To train and evaluate our models, the entire dataset was partitioned into five mutually exclusive using stratified sampling that preserved the peptide-length and HC\textsubscript{50} distribution to limit potential distribution shift. In each iteration, four folds were used to train the model and tune hyper-parameters, while the remaining fold served as an unseen validation set to prevent data leakage. Model performance is then reported as the mean $\pm$ standard deviation across the five runs.

Logarithmic transformation is widely used to stabilise variance and improve the performance of regression models \cite{changyong2014log}. Accordingly, we converted the raw HC\textsubscript{50} measurements to their negative natural logarithm (Eq.~\ref{eq:phc}) and used this value as the regression target. The transformation reduces the influence of extreme values and mitigates heteroscedasticity.

\begin{equation}
    \text{pHC}_{50} = -\ln\bigl(\text{HC}_{50}\bigr)
    \label{eq:phc} 
\end{equation}

\paragraph{Data augmentation} To combat overfitting and also encourage the model to learn contextual information from the peptide sequences, we applied random feature masking during training. For each peptide, a randomly chosen subset of 0–50 features from each per-residue protein-language-model embedding was masked. By forcing the model to reconstruct the missing information from neighboring features, this regularization promotes more context-aware representations.

\subsection{Comparison Experiments}

\begin{table*}[!t]
\centering
\begin{threeparttable}
  \caption{Comparison of performance of different models}%
  \label{table:comparison}
  \tabcolsep=2.0pt
  \begin{tabular*}{\textwidth}{@{\hskip\tabcolsep\extracolsep\fill}
      c c c c c c @{\extracolsep\fill\hskip\tabcolsep}}
    \hline
    \textbf{Model} & \textbf{Features} & \textbf{PCC} $\uparrow$ &
      \textbf{MSE} $\downarrow$ & \textbf{MAE} $\downarrow$ &
      \textbf{R$^2$} $\uparrow$ \\
    \hline
    SVR   & ALLCOMP ex SPC\tnote{a}     & 0.302 $\pm$ 0.033 & 2.152 $\pm$ 0.168 & 1.096 $\pm$ 0.044 & 0.052 $\pm$ 0.021 \\
    XGB   & ALLCOMP ex SPC              & 0.677 $\pm$ 0.017 & 1.267 $\pm$ 0.064 & 0.858 $\pm$ 0.027 & 0.441 $\pm$ 0.018 \\
    AdaBoost & ALLCOMP ex SPC           & 0.600 $\pm$ 0.027 & 1.550 $\pm$ 0.076 & 0.993 $\pm$ 0.028 & 0.316 $\pm$ 0.024 \\
    HemoPI2  & ALLCOMP ex SPC           & 0.730 $\pm$ 0.014 & 1.074 $\pm$ 0.058 & 0.753 $\pm$ 0.026 & 0.526 $\pm$ 0.019 \\[6pt]
    SVR   & ProtT5+ESM\tnote{b}         & 0.646 $\pm$ 0.027 & 1.351 $\pm$ 0.073 & 0.840 $\pm$ 0.018 & 0.404 $\pm$ 0.028 \\
    XGB   & ProtT5+ESM                  & 0.690 $\pm$ 0.022 & 1.231 $\pm$ 0.036 & 0.846 $\pm$ 0.021 & 0.456 $\pm$ 0.024 \\
    AdaBoost & ProtT5+ESM               & 0.627 $\pm$ 0.028 & 1.422 $\pm$ 0.081 & 0.943 $\pm$ 0.028 & 0.372 $\pm$ 0.032 \\
    RF    & ProtT5+ESM                  & 0.705 $\pm$ 0.025 & 1.200 $\pm$ 0.043 & 0.816 $\pm$ 0.029 & 0.470 $\pm$ 0.028 \\[6pt]
    AmpLyze & Fusion Embeddings\tnote{c}  & \textbf{0.756} $\pm$ 0.019 & \textbf{0.987} $\pm$ 0.095 & \textbf{0.703} $\pm$ 0.029 & \textbf{0.570} $\pm$ 0.033 \\
    \hline
  \end{tabular*}

  \begin{tablenotes}\footnotesize
    \item[a] ALLCOMP ex SPC: all compositional descriptors except the property-level Shannon entropy (SPC) features.  
    \item[b] ProtT5+ESM: concatenation of per-sequence embeddings from ProtT5 and mean-pooled per-residue embeddings from the ESM protein language model.  
    \item[c] Fusion Embeddings: use of per-residue embeddings from ProtT5 and ESM with the per-sequence ProtT5 embedding.
  \end{tablenotes}
\end{threeparttable}
\end{table*}

To obtain reliable estimates that are not biased by a single random split, we assessed every model under a stratified 5‑fold cross‑validation protocol. The entire dataset was first shuffled once and then partitioned into five mutually exclusive folds such that the distribution of peptide length and pHC\textsubscript{50} was preserved in each fold. Table~\ref{table:comparison} presents our results using various metrics: the mean $\pm$ standard deviation across the five test folds for Pearson’s correlation coefficient (PCC), mean squared error (MSE), mean absolute error (MAE), and the coefficient of determination ($R^2$).

We benchmarked AmpLyze against the classical regressors most frequently used in AMP property predictions—Support Vector Regressor (SVR), Random Forest (RF), AdaBoost and XGBoost—as well as the state‑of‑the‑art model HemoPI2 \cite{rathore2025prediction}. HemoPI2 is an RF model trained on the ALLCOMP ex SPC physicochemical descriptors. Among ML models trained on the same hand‑crafted features, HemoPI2 reproduced its reported strong performance, achieving a PCC of 0.705 ± 0.025 and an $R^2$ of 0.470 ± 0.028. Switching to embeddings from the large protein language models, ProtT5 and ESM2, degraded HemoPI2’s performance, consistent with observations in its original study. By contrast, it is interesting that the SVR benefited substantially from these PLM embeddings, with PCC rising from 0.302 to 0.646, suggesting that kernel‑based models are effective in high-dimensional spaces. Among these models, our model outperforms every baseline across all metrics, achieving the highest PCC (0.756 $\pm$ 0.019) and $R^2$ (0.570 $\pm$ 0.033), alongside the lowest MSE (0.987 $\pm$ 0.095) and MAE (0.703 $\pm$ 0.029). This indicates our model’s ability to capture the nonlinear relationship between the sequence and toxicity, and generalize well for more reliable results.

\begin{figure}
    \centering
    \includegraphics[width=1\linewidth]{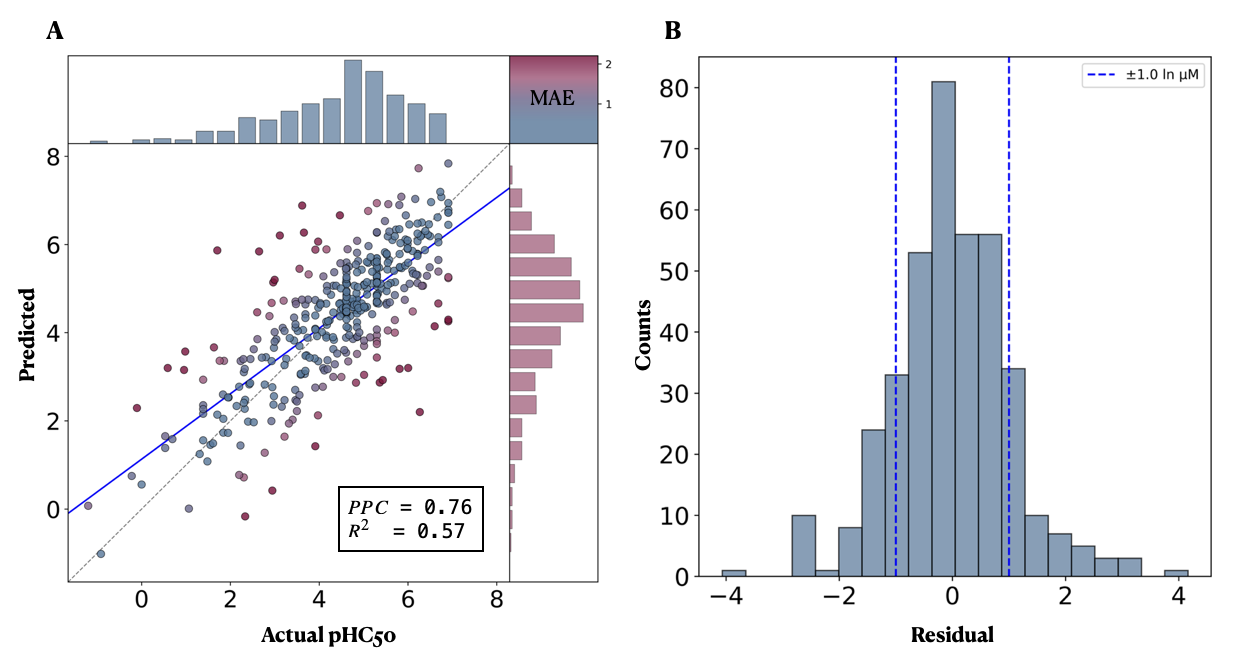}
    \caption{Performance of AmpLyze. (A) Scatter of predicted versus experimental pHC$_{50}$ values. (B) Histogram of residuals.}
    \label{fig:reg}
\end{figure}

Fig.~\ref{fig:reg}A visualizes the performance of AmpLyze on the test set comparing the predicted and experimental pHC\textsubscript{50}. The dashed blue line shows the least‑squares regression, which lies close to the identity line (grey dotted), indicating promising systematic bias. The margin histograms of the predicted (right) and observed (top) values exhibit similar shapes, indicating that  AmpLyze recovers the empirical distribution of toxicity across the range rather than concentrating the accuracy in any particular region. Scatter points colored by absolute error further show that the large deviations are well distributed. Fig.~\ref{fig:reg}B presents the distribution of residuals. Errors are tightly centered around zero and follow an approximate Gaussian distribution with light tails, with 73.3\% of predictions falling within $\pm$ 1 ln~\textmu M.

\subsection{Model application and model interpretability for per-residual importance.}

Deep learning models have been widely used for their ability to tackle various complex tasks and learn from high-dimensional data, but they also suffer from the black-box nature due to over-parameterization that impedes model interpretation \cite{interpretable}.  In recent years, extensive research has focused on illuminating these model’s decision-making processes. In our context, decoding per-residue influence on the model’s prediction of pHC\textsubscript{50} yields practical, high-resolution insights that could significantly facilitate peptide design and engineering. 

Among the methods and tools designed for model interpretability, Integrated Gradients (IG) enjoys several desirable properties such as implementation invariance, completeness and sensitivity \cite{IG}. These properties make IG the natural choice for residue-level interpretation that should capture even subtle changes in a single amino acid token in a robust manner, unaffected by model implementation while reflective of the learned biology. It attributes the prediction $F(x)$ to each input feature $x_i$ by integrating the model’s gradient over the path from the baseline $x$ to the input $x$, as described by equation~\ref{eq:IG}. Choosing the proper baseline is therefore a crucial hyperparameter. The conventional approach is to use the zero embedding vector as the baseline. However, recent work has argued that the zero vector typically lies outside the data distribution representing an impossible state and thus produces less meaningful attributions \cite{baseline}.

\begin{equation}
\mathrm{IG}_i(x) \;=\; (x_i - x_i') \int_{0}^{1}
\frac{\partial F\bigl(x' + \alpha\,(x - x')\bigr)}{\partial x_i}
\,\mathrm{d}\alpha
\label{eq:IG}
\end{equation}

\begin{figure*}
    \centering
    \includegraphics[width=0.97\linewidth]{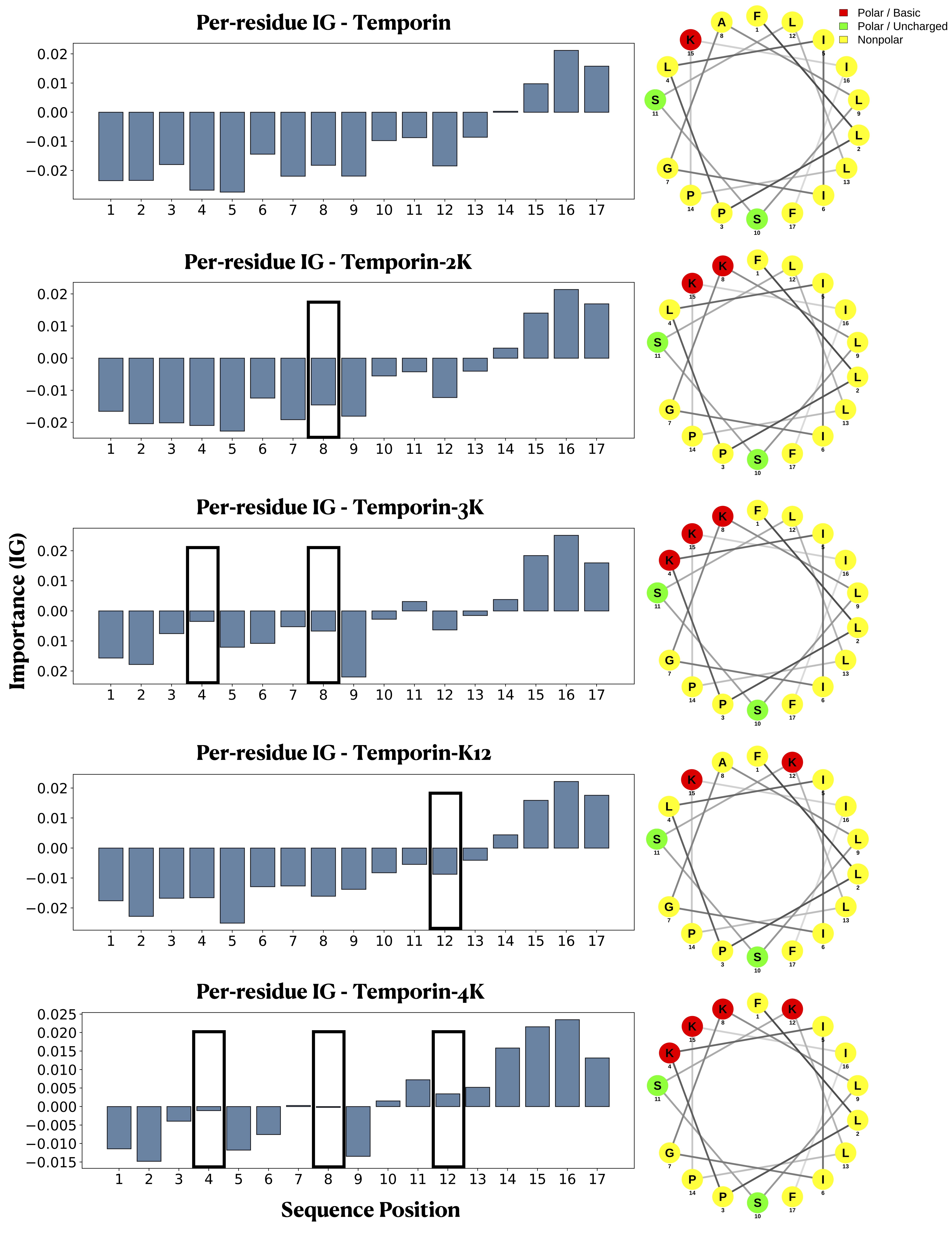}
    \caption{Per‑residue IG attributions and helical‑wheel representations for wild‑type and mutant Temporin peptides. Bar charts show per‑residue IG importance scores with higher values indiciating positive contributions to reduced hemolytic activity and lower values indicating contributions to increased hemolysis. Helical wheels color residues by polarity: red for basic/polar; green for polar/uncharged; yellow or nonpolar.}
    \label{fig:IG}
\end{figure*}

In this study, we adopted a multi‑baseline Expected Gradients (EG) strategy that replaces a single baseline with the expectation of IG over a distribution of baselines (Eq.~\ref{eq:EG}) to reduce attribution variance and sensitivity to arbitrary baseline choices \cite{erion2021improving}. We picked those peptides from the training set with pHC50 larger than 6~\textmu M. Such sequences are essentially non‑hemolytic, so they formed the reference pool that remains on‑manifold and preserves natural length and amino‑acid statistics. IG was then approximated with 50 Riemann steps and internal batch size of 32. We monitored convergence delta and discarded any attribution whose mean $|\Delta|$ exceeded 0.01 to ensure the integral faithfully reconstructs the prediction gap. Finally, per‑token attributions were averaged across the ProtT5 and ESM embedding channels and summed along the feature dimension to yield a single importance score for every residue.

\begin{equation}
\mathrm{EG}_i(x) \;=\; \mathbb{E}_{x'}
\!\Bigl[
  (x_i - x'_i)\!
  \int_{0}^{1}
  \frac{\partial F\!\bigl(x' + \alpha\,(x - x')\bigr)}{\partial x_i}
  \,\mathrm{d}\alpha
\Bigr]
\label{eq:EG}
\end{equation}

To show how interpretability can guide in silico AMP design and to further validate our model's predictive robustness, we compiled several AMP optimization examples from the literature focused on reducing hemolysis. The prediction results are summarized in Table~\ref{table:XAI}. Hemolytic activity depends on complex relationships of residue physicochemistry, peptide structures and membrane interactions, and no simple rule has been established yet for the design \cite{yeaman2003mechanisms, oddo2016hemolytic}. As a result, targeted amino acid substitutions remain the predominant optimization strategy. In this challenging setting, our model continues to perform strongly and accurately captures the effect of substitution of amino acids in most cases (PCC=0.844, MAE=0.863), reflecting that the model has indeed learned the underlying relationships. For instance, substituting two positions with lysine residues in Dermaseptin yields a dramatic increase in predicted pHC\textsubscript{50}, which is in close agreement with experimental measurements.

\begin{table*}
\centering
\begin{threeparttable}
  \caption{Prediction of hemolysis (pHC\textsubscript{50}) in parent and mutant AMPs}
  \label{table:XAI}
  \tabcolsep=1.5pt
  \begin{tabular*}{\textwidth}{@{\hskip\tabcolsep\extracolsep\fill}
      c c c c @{\extracolsep\fill\hskip\tabcolsep}}
    \hline
    \textbf{Peptide} & \textbf{Sequence} & \textbf{Predicted} &
      \textbf{Experiment} \\
    \hline
    Dermaseptin S4 \tnote{a}   &    ALWMTLLKKVLKAAAKALNAVLVGANA     & 1.049 & -0.511 \\
    Dermaseptin S4-1 \tnote{a} & ALWMTLKKKVLKAKAKALNAVLVGANA     & 4.660 & 5.485 \\
    Temporin \tnote{b}      & 
    FLPLIIGALSSLLPKIF           & 2.293 & 1.884 \\
    Temporin-2K \tnote{b}   & 
    FLPLIIGKLSSLLPKIF   & 3.865 & 2.989 \\ 
    Temporin-3K \tnote{b}             & 
    FLPKIIGKLSSLLPKIF           & 4.540 & 4.471 \\
    Temporin-K\textsubscript{12} \tnote{b}        & 
    FLPLIIGALSSKLPKIF           & 4.592 & 4.810 \\
    Temporin-4K \tnote{b}   & 
    FLPKIIGKLSSKLPKIF     & 5.210 & 7.421 \\
    CM \tnote{c}            &
    KWKLFKKIGAVLKVL     & 4.145 & 3.916 $\pm$ 0.052 \\
    CM-10K14K \tnote{c}     & 
    KWKLFKKIGKVLKKL     & 4.922 & 5.713 $\pm$ 0.157 \\
    CM-1V10K14K \tnote{c}     & 
    VWKLFKKIGKVLKKL     & 4.346 & 4.175 $\pm$ 0.069 \\
    CM-1V14K  \tnote{c}        & 
    VWKLFKKIGAVLKKL     & 3.994 & 4.310 $\pm$ 0.070 \\
    MPIII    \tnote{d}    &
    INWLKLGKAVIDAL    & 2.829 & 3.135 $\pm$ 0.052 \\
    MPIII-1  \tnote{d}    &
    INWLKLGKKVIDAL    & 4.395 & 5.173 $\pm$ 0.006 \\
    MPIII-2  \tnote{d}    &
    INWLKLGKAVSDAL    & 4.617 & 6.512 $\pm$ 0.281 \\
    MPIII-6  \tnote{d}    &
    INWLKLGKKVIAAL    & 4.598 & 4.700 $\pm$ 0.023 \\
    MPIII-8  \tnote{d}    &
    INWLKLGKAVSAAL    & 4.136 & 5.729 $\pm$ 0.020 \\
    MPIII-9  \tnote{d}    &
    INWLKLGKAVSDIL    & 3.876 & 5.120 $\pm$ 0.016 \\
    MPIII-11  \tnote{d}    &
    INWLKLGKAVSAIL    & 3.273 & 4.946 $\pm$ 0.017 \\
    MPIII-12  \tnote{d}    &
    INWLKLGKKVIAIL    & 4.930 & 3.782 $\pm$ 0.082 \\
    \hline
  \end{tabular*}

  \begin{tablenotes}\footnotesize
    \item[a] Data from \cite{peptide1}.
    \item[b] Data from \cite{peptide2}. 
    \item[c] Data from \cite{peptide3}. 
    \item[d] Data from \cite{peptide4}. 
  \end{tablenotes}
\end{threeparttable}
\end{table*}

\begin{figure}
    \centering
    \includegraphics[width=1.0\linewidth]{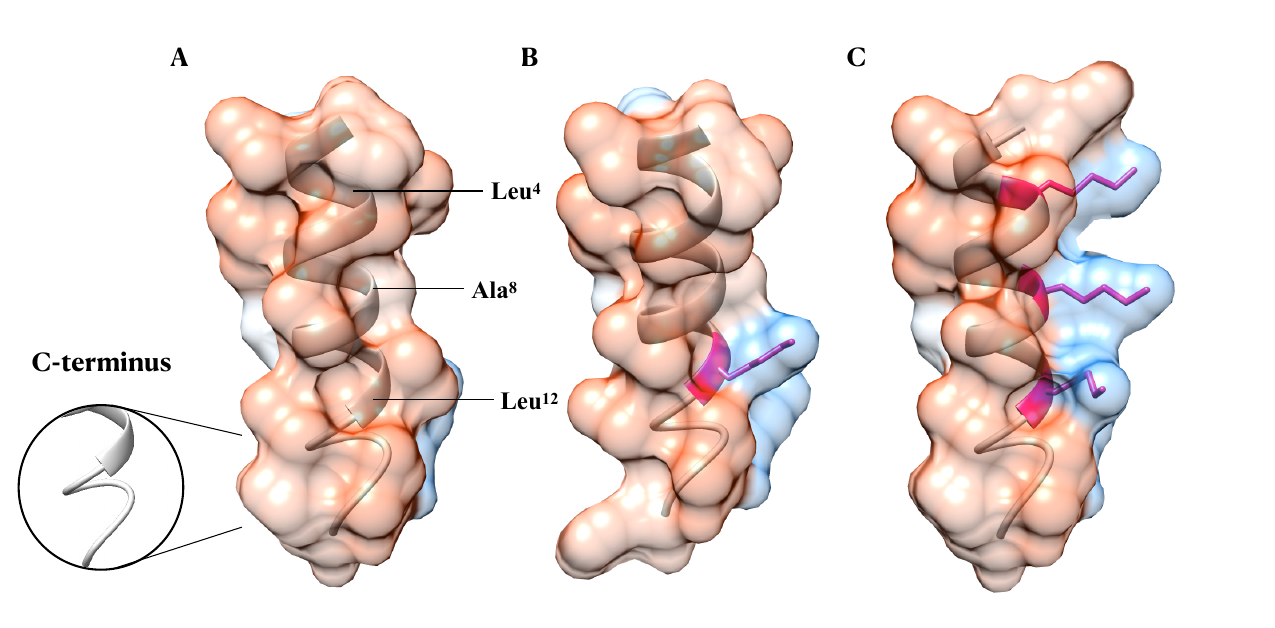}
    \caption{Surface hydrophobicity and backbone representation of wild‑type and mutant Temporin peptides. 
    (A) Wild‑type Temporin shows a continuous hydrophobicity surface with the flexible C‑terminus. 
    (B) Single mutant Temporin‑K\textsubscript{12} (substitution in magenta) disrupts the hydrophobicity surface by introducing a positively charged side chain. 
    (C) Triple mutant Temporin‑4K breaks up nonpolar clustering with three lysine substitutions. 
    Peptide 3D structures were predicted using PEP-FOLD 3.5 with default parameters \cite{peptide_structure}.}
    \label{fig:peptides}
\end{figure}

Information about the influence of each amino acid provides crucial initial optimization directions, and saves work from tedious permutations. We performed a case study on the Temporin series.  Temporins are a family of short, hydrophobic peptides secreted by frog skin, and hold high value as a template for engineering novel antimicrobial therapeutics. The IG map for the original temporin (Fig.~\ref{fig:IG}) indicates that most positions 1-13 show negative contributions, with only the tail showing a positive contribution. This corresponds to the floppy nature of the C-terminus as shown in Fig.~\ref{fig:peptides}A. Such structure usually impedes deep and stable insertion into the lipid bilayer, thereby limiting pore formation and reducing the peptide’s hemolytic activity. For its mutants, our model correctly predicts the effect of each lysine substitution, and the IG profile clearly confirms the causal link. For example, substituting Leu\textsuperscript{12} with Lys (Temporin-K\textsubscript{12}) reduces hemolytic activity into the safe range, and the IG profile indicates its negative contribution is halved, while the surrounding negatively contributing residues are damped. This reflects the mechanism whereby this added positively charged residue damages the hydrophobic face (Fig.~\ref{fig:peptides}B), with the hydrophobic moment dropping from 0.490 $\mu\mathrm{H}$ to 0.443 $\mu\mathrm{H}$. Substituting two more lysines at Leu\textsuperscript{4} and Ala\textsuperscript{8} (Temporin-4K) further perforates the non‑polar face (Fig.~\ref{fig:peptides}C). In the IG profile, the prominent negative bars that dominated the N‑terminal half of the parent peptide are greatly abolished or even flip slightly positive. Experimentally, this cumulative disruption pushes the HC\textsubscript{50} up by another order of magnitude ($> 1000~\mu\mathrm{M}$).

Hindsight is 20/20: once the data are in, one can quickly credit Temporin‑4K’s reduced hemolysis to its lower hydrophobicity. Yet making that call a priori is difficult, as physicochemical properties interact in a highly non‑additive way. Temporin 4K actually displays a higher hydrophobic moment than the wild‑type, which is normally linked to greater toxicity, while its diminished hydrophobicity drives the opposite outcome. With no universally accepted guideline for reconciling these competing factors, rational optimisation of AMPs towards lower hemolysis remains challenging. However, AmpLyze helps close this gap by offering an informative prior that could guide and accelerate AMP optimization.

\subsection{Ablation Study}

To understand the importance of each component of the architecture (Fig.~\ref{fig:model}), we performed an ablation study. After removing one module at a time, we retrained the network while keeping all other settings unchanged, such as dataset splits, learning-rate schedule and optimizer, and compared the results with the performance of the original model. The three variants are: (1) Global-only, which removes the local branch. The model bases its prediction exclusively on the per-sequence information. (2) Local-only, which strips the global branch, forcing the model to rely solely on per-residue information and ignoring the coarse sequence-level context information. (3) No Cross-attention, which retains both branches but the global-to-local cross-attention mechanism is replaced with simple mean-pooling over the local features. All ablated models were trained from scratch under identical hyper-parameter settings to ensure fair comparison. As summarized in Table~\ref{table:ablation}, dropping either representation branch degrades performance: Local-only (PCC 0.711, MSE 1.195) and Global-only (0.725, 1.107) each lose around 4-6\% PCC relative to the full model, confirming that residue‑level and sequence‑level cues are complementary. Adding cross-attention on top of the dual branches in place of simple pooling elevates performance to PCC 0.756 and MSE 0.987, providing a further 3\% reduction in MSE and 1\% gain in PCC achieved by dynamically aligning global context with critical residues.

\begin{table}
  \centering
  \begin{threeparttable}
    \caption{Performance of AmpLyze ablation variants}%
    \label{table:ablation}%
    \tabcolsep=2.0pt
    \begin{tabular*}{\columnwidth}{@{\hskip\tabcolsep\extracolsep\fill}
        >{\raggedright\arraybackslash}p{1.5cm}
        c c c c c @{\extracolsep\fill\hskip\tabcolsep}}
      \hline
      \textbf{Model} & \textbf{PCC} $\uparrow$ &
        \textbf{MSE} $\downarrow$ & \textbf{MAE} $\downarrow$ &
        \textbf{R$^2$} $\uparrow$ \\
      \hline
        $-$ Global Branch & 0.725 $\pm$ 0.026 & 1.107 $\pm$ 0.108 & 0.744 $\pm$ 0.032 & 0.518 $\pm$ 0.04 \\
       $-$ Local Branch  & 0.711 $\pm$ 0.023 & 1.195 $\pm$ 0.101 & 0.773 $\pm$ 0.028 & 0.471 $\pm$ 0.045 \\
       $-$ Cross Attention & 0.748 $\pm$ 0.021 & 1.014 $\pm$ 0.116 & 0.716 $\pm$ 0.033 & 0.550 $\pm$ 0.046 \\
      AmpLyze                       & \textbf{0.756} $\pm$ 0.019 & \textbf{0.987} $\pm$ 0.095 & \textbf{0.703} $\pm$ 0.029 & \textbf{0.570} $\pm$ 0.033 \\
      \hline
    \end{tabular*}
  \end{threeparttable}
\end{table}

\subsection{Hyperparameter Tuning}
Hyperparameters have a direct impact on model performance, so careful tuning is essential. We performed a grid search across the hyperparameter ranges listed in Table~\ref{table:HPs}. In experiments, we discovered that the loss function had the greatest influence on the results. Because the loss guides gradient descent and the learning process, it is arguably the most critical hyperparameter in deep‑learning models. In our context of hemolytic concentration, it is worth noting that HC\textsubscript{50} values are notoriously protocol‑sensitive. The same peptide, for example, can produce hemolysis ratios that vary by up to four‑fold when the assay is run on mouse versus rabbit erythrocytes, and simply substituting the positive‑control detergent can alter the value by a further approximately three‑fold \cite{diff}. Our dataset comprises HC\textsubscript{50} values measured with human, mouse, horse and rabbit erythrocytes, yet key protocol details are often left unspecified; consequently, substantial inter‑experiment variability is inevitable.  To limit overfitting and ensure generalization across these heterogeneous conditions, a robust loss function is fundamentally essential. We evaluated Mean Square Loss (MSE), Mean Absolute Loss (MAE), Huber Loss and Log-Cosh Loss (Eq.~\ref{eq:LogCosh}). As expected, MSE performed worst because of its pronounced sensitivity to outliers. Huber Loss comes with an additional hyperparameter to tune to control the switch from quadratic and linear loss. We experimented with $\delta=0.5$ and $\delta=1.0$, with both giving better results than MAE, but still slightly trailed  Log-Cosh.  Log-Cosh seamlessly integrates quadratic and linear penalties into a single smooth, twice-differentiable function, eliminating the need to hand-tune a $\delta$ hyperparameter while dampening the influence of large residuals. 

\begin{equation}
\mathcal{L}_{\mathrm{log\text{-}cosh}}(y,\hat y)
  = \log\!\bigl(\cosh(\hat y - y)\bigr)
\label{eq:LogCosh} 
\end{equation}

\begin{table}
  \centering
  \small                
  \setlength{\tabcolsep}{3pt}  
  \renewcommand{\arraystretch}{1.1} 
  \begin{threeparttable}
    \caption{Hyperparameter search space for AmpLyze.}
    \label{table:HPs}
    \begin{tabular}{l c c}
      \hline
      \textbf{Hyperparameter} & \textbf{Candidates}      & \textbf{Best} \\
      \hline
      loss function           & MAE, MSE, Huber, Log-Cosh & Log-Cosh     \\
      frequency mask          & 0, 50, 100, \dots, 300     & 100          \\
      \# heads (MSA)          & 1, 3, 5                    & 5            \\
      dropout                 & 0.1, 0.2, \dots, 0.8       & 0.3          \\
      hidden\_dim (LSTM)      & 300, 400, \dots, 800       & 500          \\
      hidden\_dim (MLP)       & 300, 400, \dots, 800       & 500          \\
      $d_{\mathrm{model}}$    & 60, 128, 200, 300          & 128
    \\
      \hline
    \end{tabular}
  \end{threeparttable}
\end{table}

\section{Conclusion}
In this work, we present AmpLyze, a sequence‑based deep learning model to predict peptide hemolytic concentration (HC\textsubscript{50}) quantitatively and interpretably. By leveraging pre-trained protein language model embeddings and a fusion‑embedding strategy, AmpLyze achieves superior performance over prior regression approaches: Pearson’s $r$ improved from 0.739 to 0.756, $R^2$ rose from 0.543 to 0.570, MSE fell from 1.074 to 0.987, and MAE declined from 0.753 to 0.703. These gains demonstrate that deep neural architectures can more effectively capture the nonlinear, high‑dimensional relationships between sequence motifs and hemolytic activity. We further show that robust loss functions are crucial for handling the inherent heterogeneity of hemolysis assays. Experimental HC\textsubscript{50} values can vary by orders of magnitude across different erythrocyte sources, buffer compositions, and protocols. We found that losses such as Log‑Cosh enables AmpLyze to generalize across noisy, multi‑source datasets. Importantly, AmpLyze offers residue‑level interpretability via Expected Gradients with a multi‑baseline scheme anchored by non‑hemolytic references. This approach successfully recapitulates literature‑reported mutational effects, shedding light on rational peptide design. 

Looking ahead, the next step is to integrate AmpLyze with in-silico MIC predictors to create a unified HC\textsubscript{50}–MIC framework. Both hemolytic and antimicrobial activities depend on peptide–membrane interactions: increased cationicity and amphipathicity promote binding to negatively charged bacterial membranes but also enhance insertion into zwitterionic erythrocyte bilayers. Future work will focus on training a multi‑task model to capture subtle sequence features that maximize bacterial pore formation while minimizing red‑cell lysis. Such a joint optimization strategy promises to accelerate the design of peptides with balanced safety and efficacy.

\bibliographystyle{IEEEtran}
\bibliography{references}

\begin{thebibliography}{10}
\providecommand{\url}[1]{#1}
\csname url@samestyle\endcsname
\providecommand{\newblock}{\relax}
\providecommand{\bibinfo}[2]{#2}
\providecommand{\BIBentrySTDinterwordspacing}{\spaceskip=0pt\relax}
\providecommand{\BIBentryALTinterwordstretchfactor}{4}
\providecommand{\BIBentryALTinterwordspacing}{\spaceskip=\fontdimen2\font plus
\BIBentryALTinterwordstretchfactor\fontdimen3\font minus \fontdimen4\font\relax}
\providecommand{\BIBforeignlanguage}[2]{{%
\expandafter\ifx\csname l@#1\endcsname\relax
\typeout{** WARNING: IEEEtran.bst: No hyphenation pattern has been}%
\typeout{** loaded for the language `#1'. Using the pattern for}%
\typeout{** the default language instead.}%
\else
\language=\csname l@#1\endcsname
\fi
#2}}
\providecommand{\BIBdecl}{\relax}
\BIBdecl

\bibitem{cdc2019antibiotic}
\BIBentryALTinterwordspacing
{Centers for Disease Control and Prevention}, ``Antibiotic resistance threats in the united states, 2019,'' U.S. Department of Health and Human Services, CDC, Atlanta, GA, Tech. Rep., 2019, cDC Stacks \#82532. [Online]. Available: \url{https://stacks.cdc.gov/view/cdc/82532}
\BIBentrySTDinterwordspacing

\bibitem{mba2022antimicrobial}
I.~E. Mba and E.~I. Nweze, ``Antimicrobial peptides therapy: an emerging alternative for treating drug-resistant bacteria,'' \emph{The Yale journal of biology and medicine}, vol.~95, no.~4, p. 445, 2022.

\bibitem{bucataru2024antimicrobial}
C.~Bucataru and C.~Ciobanasu, ``Antimicrobial peptides: Opportunities and challenges in overcoming resistance,'' \emph{Microbiological Research}, p. 127822, 2024.

\bibitem{lamb1998pexiganan}
H.~M. Lamb and L.~R. Wiseman, ``Pexiganan acetate,'' \emph{Drugs}, vol.~56, pp. 1047--1052, 1998.

\bibitem{AntiBP}
S.~Lata, B.~Sharma, and G.~P. Raghava, ``Analysis and prediction of antibacterial peptides,'' \emph{BMC bioinformatics}, vol.~8, pp. 1--10, 2007.

\bibitem{AMPScannerVr1}
D.~P. Veltri, ``A computational and statistical framework for screening novel antimicrobial peptides,'' Ph.D. dissertation, George Mason University, 2015.

\bibitem{AMPfun}
C.-R. Chung, T.-R. Kuo, L.-C. Wu, T.-Y. Lee, and J.-T. Horng, ``Characterization and identification of antimicrobial peptides with different functional activities,'' \emph{Briefings in bioinformatics}, vol.~21, no.~3, pp. 1098--1114, 2020.

\bibitem{AMPScannerVr2}
D.~Veltri, U.~Kamath, and A.~Shehu, ``Deep learning improves antimicrobial peptide recognition,'' \emph{Bioinformatics}, vol.~34, no.~16, pp. 2740--2747, 2018.

\bibitem{AI4AMP}
T.-T. Lin, L.-Y. Yang, I.-H. Lu, W.-C. Cheng, Z.-R. Hsu, S.-H. Chen, and C.-Y. Lin, ``Ai4amp: an antimicrobial peptide predictor using physicochemical property-based encoding method and deep learning,'' \emph{Msystems}, vol.~6, no.~6, pp. e00\,299--21, 2021.

\bibitem{iAMPCN}
J.~Xu, F.~Li, C.~Li, X.~Guo, C.~Landersdorfer, H.-H. Shen, A.~Y. Peleg, J.~Li, S.~Imoto, J.~Yao \emph{et~al.}, ``iampcn: a deep-learning approach for identifying antimicrobial peptides and their functional activities,'' \emph{Briefings in Bioinformatics}, vol.~24, no.~4, p. bbad240, 2023.

\bibitem{pepnet}
J.~Han, T.~Kong, and J.~Liu, ``Pepnet: an interpretable neural network for anti-inflammatory and antimicrobial peptides prediction using a pre-trained protein language model,'' \emph{Communications Biology}, vol.~7, no.~1, p. 1198, 2024.

\bibitem{AMP-Identifier}
C.~Pian, J.~Huang, Y.~Yang, Y.~Li, L.~Gao, W.~Zhao, Z.~Wang, X.~Xu, J.~Ji, Y.~Zhang \emph{et~al.}, ``Aipampds: an ai platform for antimicrobial peptide design and screening,'' \emph{bioRxiv}, pp. 2025--03, 2025.

\bibitem{yan2023deep}
J.~Yan, B.~Zhang, M.~Zhou, F.-X. Campbell-Valois, and S.~W. Siu, ``A deep learning method for predicting the minimum inhibitory concentration of antimicrobial peptides against escherichia coli using multi-branch-cnn and attention,'' \emph{Msystems}, vol.~8, no.~4, pp. e00\,345--23, 2023.

\bibitem{chung2024ensemble}
C.-R. Chung, C.-Y. Chien, Y.~Tang, L.-C. Wu, J.~B.-K. Hsu, J.-J. Lu, T.-Y. Lee, C.~Bai, and J.-T. Horng, ``An ensemble deep learning model for predicting minimum inhibitory concentrations of antimicrobial peptides against pathogenic bacteria,'' \emph{Iscience}, vol.~27, no.~9, 2024.

\bibitem{cai2025bert}
J.~Cai, J.~Yan, C.~Un, Y.~Wang, F.-X. Campbell-Valois, and S.~W. Siu, ``Bert-ampep60: A bert-based transfer learning approach to predict the minimum inhibitory concentrations of antimicrobial peptides for escherichia coli and staphylococcus aureus,'' \emph{Journal of Chemical Information and Modeling}, vol.~65, no.~7, pp. 3186--3202, 2025.

\bibitem{ruiz2014analysis}
J.~Ruiz, J.~Calderon, P.~Rond{\'o}n-Villarreal, and R.~Torres, ``Analysis of structure and hemolytic activity relationships of antimicrobial peptides (amps),'' in \emph{Advances in computational biology: proceedings of the 2nd colombian congress on computational biology and bioinformatics (CCBCOL)}.\hskip 1em plus 0.5em minus 0.4em\relax Springer, 2014, pp. 253--258.

\bibitem{plisson2020machine}
F.~Plisson, O.~Ram{\'\i}rez-S{\'a}nchez, and C.~Mart{\'\i}nez-Hern{\'a}ndez, ``Machine learning-guided discovery and design of non-hemolytic peptides,'' \emph{Scientific reports}, vol.~10, no.~1, p. 16581, 2020.

\bibitem{HemoPI}
K.~Chaudhary, R.~Kumar, S.~Singh, A.~Tuknait, A.~Gautam, D.~Mathur, P.~Anand, G.~C. Varshney, and G.~P. Raghava, ``A web server and mobile app for computing hemolytic potency of peptides,'' \emph{Scientific reports}, vol.~6, no.~1, p. 22843, 2016.

\bibitem{HemoPred}
T.~S. Win, A.~A. Malik, V.~Prachayasittikul, J.~E. S~Wikberg, C.~Nantasenamat, and W.~Shoombuatong, ``Hemopred: a web server for predicting the hemolytic activity of peptides,'' \emph{Future medicinal chemistry}, vol.~9, no.~3, pp. 275--291, 2017.

\bibitem{AMPDeep}
M.~Salem, A.~Keshavarzi~Arshadi, and J.~S. Yuan, ``Ampdeep: hemolytic activity prediction of antimicrobial peptides using transfer learning,'' \emph{BMC bioinformatics}, vol.~23, no.~1, p. 389, 2022.

\bibitem{ToxIBTL}
L.~Wei, X.~Ye, T.~Sakurai, Z.~Mu, and L.~Wei, ``Toxibtl: prediction of peptide toxicity based on information bottleneck and transfer learning,'' \emph{Bioinformatics}, vol.~38, no.~6, pp. 1514--1524, 2022.

\bibitem{tAMPer}
H.~Ebrahimikondori, D.~Sutherland, A.~Yanai, A.~Richter, A.~Salehi, C.~Li, L.~Coombe, M.~Kotkoff, R.~L. Warren, and I.~Birol, ``Structure-aware deep learning model for peptide toxicity prediction,'' \emph{Protein Science}, vol.~33, no.~7, p. e5076, 2024.

\bibitem{rathore2025prediction}
A.~S. Rathore, N.~Kumar, S.~Choudhury, N.~K. Mehta, and G.~P. Raghava, ``Prediction of hemolytic peptides and their hemolytic concentration,'' \emph{Communications Biology}, vol.~8, no.~1, p. 176, 2025.

\bibitem{diff}
I.~P. S{\ae}b{\o}, M.~Bj{\o}r{\aa}s, H.~Franzyk, E.~Helgesen, and J.~A. Booth, ``Optimization of the hemolysis assay for the assessment of cytotoxicity,'' \emph{International journal of molecular sciences}, vol.~24, no.~3, p. 2914, 2023.

\bibitem{greco2020correlation}
I.~Greco, N.~Molchanova, E.~Holmedal, H.~Jenssen, B.~D. Hummel, J.~L. Watts, J.~H{\aa}kansson, P.~R. Hansen, and J.~Svenson, ``Correlation between hemolytic activity, cytotoxicity and systemic in vivo toxicity of synthetic antimicrobial peptides,'' \emph{Scientific reports}, vol.~10, no.~1, p. 13206, 2020.

\bibitem{kumar2016single}
A.~Kumar, A.~K. Tripathi, M.~Kathuria, S.~Shree, J.~K. Tripathi, R.~Purshottam, R.~Ramachandran, K.~Mitra, and J.~K. Ghosh, ``Single amino acid substitutions at specific positions of the heptad repeat sequence of piscidin-1 yielded novel analogs that show low cytotoxicity and in vitro and in vivo antiendotoxin activity,'' \emph{Antimicrobial Agents and Chemotherapy}, vol.~60, no.~6, pp. 3687--3699, 2016.

\bibitem{pirtskhalava2021physicochemical}
M.~Pirtskhalava, B.~Vishnepolsky, M.~Grigolava, and G.~Managadze, ``Physicochemical features and peculiarities of interaction of amp with the membrane,'' \emph{Pharmaceuticals}, vol.~14, no.~5, p. 471, 2021.

\bibitem{ESM2}
Z.~Lin, H.~Akin, R.~Rao, B.~Hie, Z.~Zhu, W.~Lu, N.~Smetanin, A.~dos Santos~Costa, M.~Fazel-Zarandi, T.~Sercu, S.~Candido \emph{et~al.}, ``Language models of protein sequences at the scale of evolution enable accurate structure prediction,'' \emph{bioRxiv}, 2022.

\bibitem{ProtT5}
\BIBentryALTinterwordspacing
A.~Elnaggar, M.~Heinzinger, C.~Dallago, G.~Rehawi, Y.~Wang, L.~Jones, T.~Gibbs, T.~Feher, C.~Angerer, M.~Steinegger, D.~BHOWMIK, and B.~Rost, ``Prottrans: Towards cracking the language of life{\textquoteright}s code through self-supervised deep learning and high performance computing,'' \emph{bioRxiv}, 2020. [Online]. Available: \url{https://www.biorxiv.org/content/early/2020/07/21/2020.07.12.199554}
\BIBentrySTDinterwordspacing

\bibitem{hemolytik}
A.~Gautam, K.~Chaudhary, S.~Singh, A.~Joshi, P.~Anand, A.~Tuknait, D.~Mathur, G.~C. Varshney, and G.~P. Raghava, ``Hemolytik: a database of experimentally determined hemolytic and non-hemolytic peptides,'' \emph{Nucleic acids research}, vol.~42, no.~D1, pp. D444--D449, 2014.

\bibitem{dbaasp}
M.~Pirtskhalava, A.~A. Amstrong, M.~Grigolava, M.~Chubinidze, E.~Alimbarashvili, B.~Vishnepolsky, A.~Gabrielian, A.~Rosenthal, D.~E. Hurt, and M.~Tartakovsky, ``Dbaasp v3: database of antimicrobial/cytotoxic activity and structure of peptides as a resource for development of new therapeutics,'' \emph{Nucleic acids research}, vol.~49, no.~D1, pp. D288--D297, 2021.

\bibitem{changyong2014log}
F.~Changyong, W.~Hongyue, L.~Naiji, C.~Tian, H.~Hua, L.~Ying, and M.~T. Xin, ``Log-transformation and its implications for data analysis,'' \emph{Shanghai archives of psychiatry}, vol.~26, no.~2, p. 105, 2014.

\bibitem{interpretable}
X.~Li, H.~Xiong, X.~Li, X.~Wu, X.~Zhang, J.~Liu, J.~Bian, and D.~Dou, ``Interpretable deep learning: Interpretation, interpretability, trustworthiness, and beyond,'' \emph{Knowledge and Information Systems}, vol.~64, no.~12, pp. 3197--3234, 2022.

\bibitem{IG}
M.~Sundararajan, A.~Taly, and Q.~Yan, ``Axiomatic attribution for deep networks,'' in \emph{International conference on machine learning}.\hskip 1em plus 0.5em minus 0.4em\relax PMLR, 2017, pp. 3319--3328.

\bibitem{baseline}
J.~Bardhan, C.~Neeraj, M.~Rawat, and S.~Mitra, ``Constructing sensible baselines for integrated gradients,'' \emph{arXiv preprint arXiv:2412.13864}, 2024.

\bibitem{erion2021improving}
G.~Erion, J.~D. Janizek, P.~Sturmfels, S.~M. Lundberg, and S.-I. Lee, ``Improving performance of deep learning models with axiomatic attribution priors and expected gradients,'' \emph{Nature machine intelligence}, vol.~3, no.~7, pp. 620--631, 2021.

\bibitem{yeaman2003mechanisms}
M.~R. Yeaman and N.~Y. Yount, ``Mechanisms of antimicrobial peptide action and resistance,'' \emph{Pharmacological reviews}, vol.~55, no.~1, pp. 27--55, 2003.

\bibitem{oddo2016hemolytic}
A.~Oddo and P.~R. Hansen, ``Hemolytic activity of antimicrobial peptides,'' in \emph{Antimicrobial peptides: methods and protocols}.\hskip 1em plus 0.5em minus 0.4em\relax Springer, 2016, pp. 427--435.

\bibitem{peptide1}
Z.~Jiang, L.~Gera, C.~T. Mant, and R.~S. Hodges, ``Design of new antimicrobial peptides (amps) with “specificity determinants” that encode selectivity for gram negative pathogens and remove both gram-positive activity and hemolytic activity from broad-spectrum amps,'' in \emph{Proceedings of the 24th American Peptide Symposium}, 2015, pp. 245--248.

\bibitem{peptide2}
Y.~Lin, Y.~Jiang, Z.~Zhao, Y.~Lu, X.~Xi, C.~Ma, X.~Chen, M.~Zhou, T.~Chen, C.~Shaw \emph{et~al.}, ``Discovery of a novel antimicrobial peptide, temporin-pke, from the skin secretion of pelophylax kl. esculentus, and evaluation of its structure-activity relationships,'' \emph{Biomolecules}, vol.~12, no.~6, p. 759, 2022.

\bibitem{peptide3}
N.~Klubthawee, M.~Wongchai, and R.~Aunpad, ``The bactericidal and antibiofilm effects of a lysine-substituted hybrid peptide, cm-10k14k, on biofilm-forming staphylococcus epidermidis,'' \emph{Scientific Reports}, vol.~13, no.~1, p. 22262, 2023.

\bibitem{peptide4}
X.~Ye, H.~Zhang, X.~Luo, F.~Huang, F.~Sun, L.~Zhou, C.~Qin, L.~Ding, H.~Zhou, X.~Liu \emph{et~al.}, ``Characterization of the hemolytic activity of mastoparan family peptides from wasp venoms,'' \emph{Toxins}, vol.~15, no.~10, p. 591, 2023.

\bibitem{peptide_structure}
A.~Lamiable, P.~Th{\'e}venet, J.~Rey, M.~Vavrusa, P.~Derreumaux, and P.~Tuff{\'e}ry, ``Pep-fold3: faster de novo structure prediction for linear peptides in solution and in complex,'' \emph{Nucleic acids research}, vol.~44, no.~W1, pp. W449--W454, 2016.

\end{thebibliography}

\end{document}